\documentclass [11pt]{article}
\usepackage[dvips]{color}
\usepackage{epsfig}
\usepackage[normalem]{ulem}

\def\beq{\begin{equation}}
\def\eeq{\end{equation}}
\def\beqa{\begin{eqnarray}}
\def\eeqa{\end{eqnarray}}
\def\d{{\rm d}}

\begin{document}
\baselineskip0.7cm plus 1pt minus 1pt
\tolerance=1500

\begin{center}
{\LARGE\bf On the work of internal forces}
\vskip0.4cm
{ J. G\"u\'emez$^{a,}$\footnote{guemezj@unican.es},
M. Fiolhais$^{b,}$\footnote{tmanuel@teor.fis.uc.pt.},
L. Brito$^{b,}$\footnote{lucilia@teor.fis.uc.pt.},
}
\vskip0.1cm
{\it $^a$ Departamento de F\'{\i}sica Aplicada}\\ {\it Universidad de
Cantabria} \\ {\it E-39005 Santander, Spain} \\
\vskip0.1cm
{\it $^b$ Departamento de F\'\i sica and Centro de
F\'\i sica Computacional}
\\ {\it Universidade de Coimbra}
\\ {\it P-3004-516 Coimbra, Portugal}
\end{center}

\begin{abstract}
We discuss the role of the internal forces and how their work changes the energy of a system. We illustrate the contribution of the internal work to the variation of the
system's energy, using a pure mechanical example, a thermodynamical system and an example from electromagnetism. We emphasize that internal energy variations related to the work of the internal forces
should be pinpointed in the classroom and placed on the same footing as other internal energy variations such as those caused by temperature changes or by chemical reactions.
\end{abstract}

\section{Introduction}
\label{sec:intro}

Discussions on internal forces reduce sometimes to the statement that their resultant, according to Newton's third law,
always vanishes. However, in general, the work of these forces is not zero and  such
work provides the explanation for internal energy variations (a concept usually associated only with thermodynamics, but of  general applicability). This paper is about the important role played by the internal forces, in particular their work, in mechanics, electromagnetism and thermodynamics. This interesting topic is probably not so well mastered by instructors and it is even a source of some misunderstandings, as pointed out in~\cite{papersAJP}.

The first law of thermodynamics is a statement on the energy conservation. Once the system is defined, it is delimited by a boundary, actual or notional, that keeps it separated from the surrounding (the part of the rest of the universe that may interact with the system). The first law of thermodynamics simply states that the variation of the total energy of the system, $\Delta E_{\rm total}$, equals the energy that crosses the system's boundary.

In recent papers \cite{guemez13, guemez14}, we proposed to split the total energy of the system, viewing it as the sum of two terms: the kinetic energy of the centre of mass of the system, $K_{\rm cm}$, and the remaining energy, identified
as the internal energy of the system, $U$. This internal energy is, therefore, the energy of the system that is not the centre of mass kinetic energy, hence $E_{\rm total}= K_{\rm cm}+U$. The first law of thermodynamics is then
expressed in the following form:
\beq
\Delta {K}_{\rm cm} + \Delta { U} = W_{\rm ext}  \ + \ Q\, .
\label{totale}
\eeq
The left hand side accounts for the system total energy variation and the right hand side refers to the energy that crosses the system's boundary. On the one hand, this energy transfer is work performed by the external forces, $W_{\rm ext}$. Indeed, any external force acting upon the system, in general gives rise to an energy transfer to/from the system, and the total external work is given by
\beq
W_{\rm ext} = {\color{black} \sum_{j=1}^N }\int { {\vec F}^{\rm ext}_j}\cdot \ \d {\vec r}_j\, ,  \label{trabalhoe}
\eeq
where ${\vec F}^{\rm ext}_j$ represents each external force acting on the system, and $\d {\vec r}_j$ its own infinitesimal displacement. {\color{black} In (\ref{trabalhoe}) we assume $N$ external forces acting upon the system and $j$ is used to label these external forces.}
On the other hand, any other energy transfer that cannot be associated with an external force that displaces its application point, should be included as heat, $Q$, in equation (\ref{trabalhoe}). In principle, there is no need to split $Q$ into several terms, or even to write $W_{\rm ext}   +  Q$ as a sum with much more terms, but it is legitimate to do so as some authors actually do \cite{afinn}, provided all the energy transfers are taken into account. In the way equation (\ref{totale}) is written, heat is  the energy crossing the system boundary that cannot be assigned to external work.
{\color{black} Of course, we may look at the energetic of a system from two different viewpoints: either through the energy that crosses the system boundary or through the consequences inside the system. These two perspectives are distinct but they are equivalent, and the internal work plays a central role in the latter.}

Once the energy had crossed the system boundary, it may either change the kinetic energy of the system centre of mass --- the first term on the left hand side of (\ref{totale}) ---, or the internal energy of the system. Of course, equation (\ref{totale}) also allows for a direct conversion of $U$ into ${K}_{\rm cm}$ and vice-versa, even when there is no energy crossing the system's boundary.

 The splitting of the internal energy variation as $\Delta U=\Delta U_{\rm k}+\Delta U_{\rm int}$, where the first term is the variation of the kinetic energy of the system in the centre of mass reference frame, and $\Delta U_{\rm int}$ is their interaction energy variation, is general, common and sometimes useful. However, it is also usual and useful to split the various contributions to $\Delta U$ in different ways, as one usually does in thermodynamics \cite{zemansky}.
 In this way, internal energy variations are regarded,  not as resulting from an energetic system/surrounding interaction, but rather as a consequence of changes in the state variables, as we shall exemplify along the paper.
 For instance, temperature variations lead, in general, to a variation of the internal energy (here denoted by $\Delta U_T$); chemical reactions also lead to variations of the internal energy (here denoted by $\Delta U_\xi$). As noticed above, the kinetic energies, both translational or rotational, with respect to the centre of mass of the system, also contribute to the internal energy, and the corresponding variation is denoted here by $\Delta U_{\rm k}$. The expression that summarizes the  various terms referred to so far, as the different contributions to the total internal energy, is
 \beq
\Delta U = \Delta U_T + \Delta U_\xi + \Delta U_{\rm k}+... \, ,
\label{decomp}
\eeq
but it is not an exhaustive one. {\color{black} In particular, it may even not be possible to look at $\Delta U$ as a sum of independent terms. In that case, we may still write down an equation like  (\ref{decomp})  that now includes ``crossed" terms. However, in the examples considered in this work, equation  (\ref{decomp}) applies}. One term that is still clearly missing refers to the possible work of internal forces, $\Delta U_{\rm w}$, whose contribution to the total internal energy variation amounts to  $-W_{\rm int}$. The reason for the minus sign is clear: by writing the first law as in equation (\ref{totale}) we adopted the convention that positive external work (or heat) increases the system's energy. Now, the internal work contribution, which is work anyway, enters on the left hand side of the equation, so  the sign should be reversed. Physically this means that a negative (positive) work performed by internal forces leads to a system's internal energy increase (decrease). {\color{black} We stress that the internal work is energy that does not cross the system boundary, hence it has to be assigned to an internal energy variation.}
However, sometimes the internal work just intermediates a conversion of a certain ``type" of internal energy variation, $\Delta U_i$, into other ``type", $\Delta U_j$, therefore it is advisable not to always explicitly include it in (\ref{decomp}), otherwise we may incur in double counting.

The main goal of this paper is precisely to stress the role of the work of the internal forces in the variation of the internal energy, $\Delta U_{\rm w} = -W_{\rm int}$, using common examples where the work of the internal forces explicitly shows up as  internal energy variations.
In some cases, $\Delta U_{\rm w}$ can be ultimately expressed in terms of  state variables and their variations but, in general, the explicit
{\em ab initio}
calculation of $W_{\rm int}$, using an expression similar to
(\ref{trabalhoe}) now for the internal forces,
\beq
W_{\rm int} = \sum_j \int { {\vec F}^{\rm int}_j}\cdot \ \d {\vec r}_j\, ,  \label{trabalhoi}
\eeq
is sometimes possible in mechanics but usually not possible in thermodynamics. However, if one knows the total variation, $\Delta U$, and if one knows all the contributions except the one corresponding to the work of the internal forces, one may express $W_{\rm int}$ in terms of variations of state variables, for instance, volume variations. We illustrate this point with the van der Waals gas. Besides pure thermodynamical examples we show, in the next section, the role of the work of the internal forces in a pure mechanical example and, in section 4, we analyze an example taken from electromagnetism.

\section{Work of the internal forces --- a pure mechanical example}

We first discuss the role of the internal forces in mechanics. To make the discussion more concrete and assertive we use the simple example of two gravitationally interacting point-like objects.
Of course this is a pure mechanical system but equation (\ref{totale}) does apply (it applies to all systems!) with $Q=0$ and interpreting the internal energy of the system as its total energy except the energy of the centre of mass that is explicitly taken into account by the first term on the left hand side of that equation.

At this point we should mention Newton's second law, which is independent from the first law of thermodynamics, and does also apply to any system. After applying the third law, ${\vec F}_{\rm int}=\sum_j  {\vec F}^{\rm int}_j=\vec 0$, Newton's second law, can be expressed in the following scalar form {\color{black} (usually referred to as work-energy theorem)}:
\beq
\Delta K_{\rm cm} = \int  {\vec F}_{\rm ext} \cdot \d {\vec r}_{\rm cm}\,
\label{eq-3}
\eeq
where ${\vec F}_{\rm ext}=\sum_j  {\vec F}^{\rm ext}_j$
is the resultant of the external forces and $\d \vec r_{\rm cm}$ is the infinitesimal displacement of the centre of mass.
We stress that the right hand side in (\ref{eq-3}) is not the external work given by equation (\ref{trabalhoe}) and, therefore, both expressions should not be confused. In
(\ref{trabalhoe}) each external force and its own displacement is considered separately, whereas in  (\ref{eq-3}) is is the resultant of the external forces and the displacement of the centre of mass that appears. To emphasize the distinction, the right hand side of (\ref{eq-3}) is usually referred to as pseudowork, but the name does not matter. The important point is to realize that, in general, $W_{\rm ext}\not= \Delta K _{\rm cm}$ \cite{penchina78,sherwood83}.

\begin{figure}[htb]
\begin{center}
\hspace*{-0.5cm}
\includegraphics[width=9.5cm]{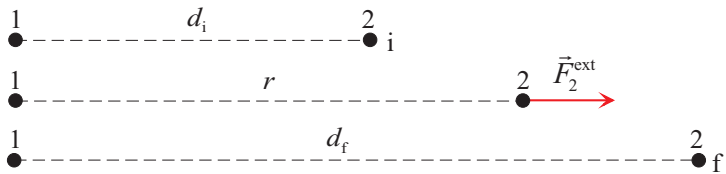}
\end{center}
\vspace*{-0.5cm}
\caption[]{\label{fig1} \small Particle 2 is acted upon by an external force and getting away from particle 1 (which is fixed).
} %\vspace*{-0.2cm}
\end{figure}

Our mechanical example consists of two particles with masses, $m_1$ and $m_2$, initially at rest and separated by a distance $d_{\rm i}$.
With the particles at rest, the total energy of the system is {\color{black} solely the internal energy} given by
\beq
U= -G\frac{m_1 m_2}{d_{\rm i}}\, .
\label{potm}
\eeq
Let us assume that particle 1 is fixed but particle 2 may be moved from its initial position, i, to a final position, f. The initial and final separations of the two particles are
$d_{\rm i}$ and $d_{\rm f}>d_{\rm i}$. Particle 2 moves due to the action of an external force, $\vec F^{\rm ext}_2$, as shown in figure \ref{fig1}.

\begin{figure}[htb]
\begin{center}
\hspace*{-0.5cm}
\includegraphics[width=9.5cm]{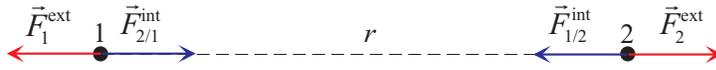}
\end{center}
\vspace*{-0.5cm}
\caption[]{\label{fig2} \small External and internal forces acting on the particles at a generic position in the process.
} %\vspace*{-0.2cm}
\end{figure}

The forces acting on the particles are: on particle 2, the external force,  $\vec F_{2}^{\rm ext}$, which displaces its application point from left to right, and the attractive internal force,   $\vec F_{1/2}^{\rm int}$, due to particle 1; on particle 1, the attractive internal force,   $\vec F_{2/1}^{\rm int}$, due to particle 2 and an external force, $\vec F_{1}^{\rm ext}=-\vec F_{2/1}^{\rm int}$ --- note that these two forces are responsible for keeping particle 1 at the same position, so they do not displace their application points. The two internal forces form an action-reaction pair. At a generic intermediate position during the process the four forces acting on the particles are represented in figure \ref{fig2}.
Particle 2 may be moved with an arbitrary (even varying) velocity along its trajectory, but we impose zero initial and final velocities.  For instance, we may have $\vec F_{2}^{\rm ext}=-\vec F_{1/2}^{\rm int} $ except for an initial short time interval and a final equally short time interval. During these time intervals, the external force is slightly larger and smaller than the internal force, respectively. In these way, particle 2 is moved with constant small velocity (this mental exercise is usually referred to as the virtual work method \cite{ray06}).

There is no variation of the centre of mass kinetic energy, hence equation (\ref{eq-3}) must vanish:
\beq
\Delta K_{\rm cm} = \int  \left( {\vec F}_1^{\rm ext} +  {\vec F}_2^{\rm ext}  \right)  \cdot \d {\vec r}_{\rm cm} = 0
\label{eq-4}
\eeq
and, since the centre of mass displaces as particle 2 moves, the equation is compatible with ${\vec F}_{1}^{\rm ext} =-  {\vec F}_{2}^{\rm ext}$ (of course, these forces do not form an action-reaction pair).

From the  point of view of equation (\ref{totale}), since {\color{black} the initial and final kinetic energies are equal,} $\Delta K_{\rm cm}=0$, and also $Q=0$, {\color{black} and that} equation reduces to
\beq
\Delta U = W_{\rm ext}\, .
\label{eqdas}
\eeq
This is not a surprising result since it means that the (positive) external work performed upon the system results in an increment of its internal energy. The external work (work of the external forces) is
\beq
W_{\rm ext} = \int  \left( {\vec F}_{1}^{\rm ext}  \cdot \d {\vec r}_1  +  {\vec F}_{2}^{\rm ext} \cdot \d {\vec r}_2  \right) \, .
\label{eq-5}
\eeq

The first term in the integral vanishes, because particle 1 does not move. Hence one has $W_{\rm ext}=W^{\rm ext}_2$ and the integral  is readily evaluated, yielding (see equation (\ref{potm}))
\beq
W_{\rm ext} = -G m_1 m_2 \left( {1 \over d_{\rm f}} - {1 \over d_{\rm i}} \right) = \Delta U > 0.
\label{eq-6}
\eeq

As we mentioned in the Introduction, the work of the internal forces leads to a variation of the internal energy which is the symmetric of that work. Let's see explicitly that this is observed in the present situation. The internal force on particle 1 does not perform any work because it does not displace its application point, therefore the internal work reduces to the work of the internal force on particle 2: $W_{\rm int}=W^{\rm int}_2$. Concerning the work of the internal force on particle 2, it is exactly the symmetric of the work of the external force acting on that particle, $W^{\rm int}_2=-W^{\rm ext}_2$. Hence, using equation (\ref{eqdas})
\beq
\Delta U = - W_{\rm int} \, .
\label{eqdasi}
\eeq
Equations (\ref{eqdas}) and (\ref{eqdasi}) express two different points of view. One, consists in  looking at the energy transfer that flows through the boundary and the concomitant variation of the internal energy --- this is the point of view of equation (\ref{eqdas}). The other point of view consists in
thinking exclusively in the system, and not on the energy transfer that flows through the boundary. In this case one may wonder about the origin of the internal energy variation in terms of phenomena that have happened inside the system. The answer is clear: the increase of the internal energy is due to the work of the internal forces {\color{black} and it is the negative of this work} --- this is the alternative point of view expressed by equation (\ref{eqdasi}). Though the sum of the internal forces vanishes, the work of these forces does not vanish, leading to a system's energy variation, and the present example explicitly illustrates  the situation.

{\color{black} The mechanical example studied in this section is rather simple and we may argue that a reference to the role of the work of the internal forces is not necessary to better understand the given situation. However, for the examples studied in the next sections, the role played by the internal forces is less trivial though  it is definitely very helpful for a better understanding of the energetic of the system. The pedagogical discussion carried on in this section, for the case of a simple example, is useful to better understand the forthcoming more complicated situations.}

\section{Thermodynamical example}

\begin{figure}[htb]
\begin{center}
\hspace*{-0.5cm}
\includegraphics[width=12cm]{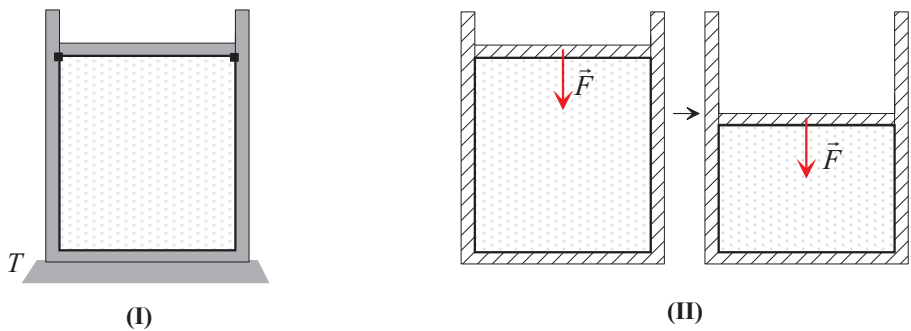}
\end{center}
\vspace*{-0.5cm}
\caption[]{\label{figt} \small Gas inside a cylinder  of rigid walls in diathermic contact with a heat reservoir (I);  cylinder surrounded by an adiabatic boundary  with a moveable piston where an external force is applied (II).
} %\vspace*{-0.2cm}
\end{figure}

Let us now analyze thermodynamical examples and, to this end, we consider an ideal gas confined by a piston  to a cylinder at rest. Firstly, we assume that the piston is fixed, so there is no external work upon the system. Any energetic interaction takes place through a heat transfer (see figure \ref{figt}-I). The amount of energy transferred, $Q$, leads to a variation of the internal energy, according to equation (\ref{totale}), given by
\beq
Q = \Delta U = \Delta U_T = n c_v \Delta T \, ,
\label{idealg2}
\eeq
where $n$ is the amount of matter in the system (in moles), $c_v$ is the molar specific heat and $\Delta T$ is the temperature variation. Now, if the cylinder (including the piston) is coated by an adiabatic boundary, the energy transfer to the system is still possible through external work. One may, for instance, reduce the system volume by applying an external force to the piston as shown in figure \ref{figt}-II.   Equation (\ref{totale}) now yields
\beq
W_{\rm ext}= \Delta U = \Delta U_T = n c_v \Delta T
\label{idealg3}
\eeq
Moreover, if the heat in situation I equals the external work in situation II the increase of internal energy is the same in both cases, i.e. the final temperatures will be the same if they were initially the same.

In either case, we may also take the point of view of the system's ``inside" and wonder why the system's internal energy has changed in the processes. The answer is the same in both situations:  the temperature has increased.
This explanation is alternative to the other --- the one that says that the energy variation is the result of an energy transfer to the system through its boundary. Both explanations are correct, and address complementary perspectives: the former, the point of view centred exclusively in the system; the latter, the point of view based on the energetic interaction system/surrounding.

For the ideal gas, the variation of the internal energy is not due to internal work. In the microscopic view of the ideal gas, its constituents are point like objects that do not exert any force upon each other, except when they collide elastically. Therefore there cannot be any work performed by internal forces. Hence, the total internal energy variation is solely due to the
temperature variation, as expressed by equations (\ref{idealg2}) and (\ref{idealg3}).

The situation is different for a real gas, for instance a gas described by the van der Waals equation of state. The energetic equation for the substance is given by \cite{zemansky}
\beq
\Delta U = n c_v (T_{\rm f} - T_{\rm i}) - n^2 a \left( {1 \over V_{\rm f}}- {1 \over V_{\rm i}}  \right)
\label{vdwe}
\eeq
where $a$ is a parameter characterizing the gas and $(T_{\rm i},V_{\rm i})$ and $(T_{\rm f},V_{\rm f})$ are the initial and final temperatures and volumes. For situation I in figure \ref{figt}, an expression similar to (\ref{idealg2}) also applies to the van der Waals gas, so there is no work of the internal forces. However, for situation II, the external work,  $W_{\rm ext}$, performed upon the system due to the external force $\vec F$, is equal to the total internal energy variation, as in the first equality in (\ref{idealg3}), but now we may write the internal energy variation as
\beq
\Delta U = \Delta U_T + \Delta U_{\rm w} \ \ \ \ \ \ \ {\rm with} \ \ \ \left\{
\begin{array}{l}
\Delta U_T = n c_v (T_{\rm f} - T_{\rm i}) \\
\Delta U_{\rm w} = - n^2 a \left( {1 \over V_{\rm f}}- {1 \over V_{\rm i}} \right).
\end{array}
\right.
\eeq
The last term stands for the work of the internal forces, $\Delta U_{\rm w}=-W_{\rm int}$.
If $V_{\rm f}<V_{\rm i}$, the work  $W_{\rm int}=n^2 \, a \left( {1 \over V_{\rm f}}- {1 \over V_{\rm i}} \right)$
is positive,  therefore responsible for a decrease of the internal energy of the system. When  the energy that crosses the system boundary (heat or work) is the same for situations I and II in figure \ref{figt}, the final temperature in II will be certainly higher than in I.

For an isothermal compression of the van der Waals gas, $\Delta U<0$, which means that the heat transfer to the surrounding is larger than the external work performed upon the system. Alternatively, one may also argue that there was positive work done by the internal forces, causing the internal energy reduction.

Contrary to the case studied in section 2, now one  cannot explicitly obtain the internal work from an {\em ab initio} calculation, using equation (\ref{trabalhoi}). This work is a sum over a myriad of ``microscopic" (elementary) works performed by each internal force acting on each particle.
It is not even possible to represent graphically the internal forces as we have done for the particles in figure \ref{fig2}. Notwithstanding, the internal work does exist and it is responsible for an energy variation experienced by the system.

{\color{black} In the case of the ideal gas mentioned at the beginning of the section, $W_{\rm int}=0$ so the work of the internal forces doesn't even exist. However, the internal work is a usefull concept if we want to understand the energetic of more sophisticated systems, such as equation (\ref{vdwe}), for the wan der Waals fluid, in the spirit of equation (\ref{decomp}).}

\section{Capacitor at constant potential}

\begin{figure}[htb]
\begin{center}
\hspace*{-0.5cm}
\includegraphics[width=15cm]{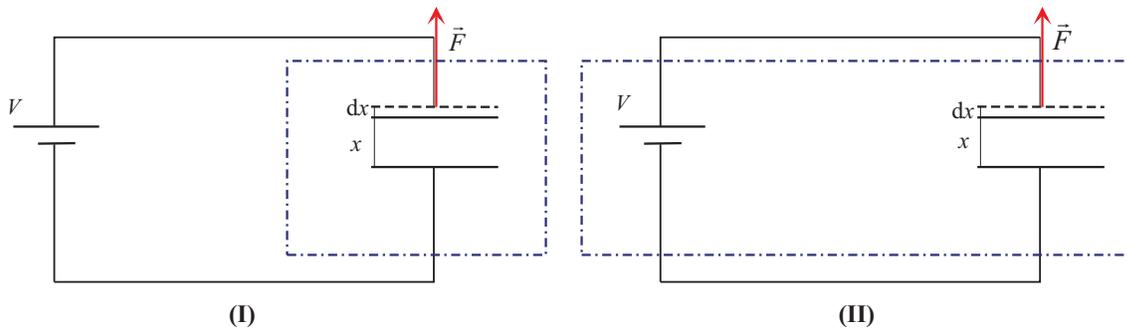}
\end{center}
\vspace*{-0.5cm}
\caption[]{\label{fig4n} \small External force acting upon the upper plate of a capacitor kept at a constant potential. In (I) the system is the capacitor. In (II) the system also includes the battery.
} %\vspace*{-0.2cm}
\end{figure}

In this section we revisit the well known example (see e.g. \cite{wangsness86}) of a capacitor of variable capacity connected to an ideal battery that keeps it at a constant potential.
We consider a parallel plate capacitor with variable capacity due to the possibility of changing the distance between the two plates.
The capacity is given by $C=\epsilon_0 {A\over x}$ where $\epsilon_0$ is the vacuum electric permittivity, $A$ the area of the planar plate
and $x$ the variable distance between the two plates, as shown in figure \ref{fig4n}.

For the discussion, it is important to define the thermodynamical system. We consider two possibilities: I --- the system is solely the capacitor; II --- the system also includes the battery. Once the system is defined, one has to apply  expression (\ref{totale}) accordingly.

Let us first consider case I.
The work performed by the external force, represented in figure \ref{fig4n}, is $\delta W = F \d x$ for an infinitesimal displacement of the upper plate that moves upwards. Moreover, there is work of the electric forces, whose origin lies in the battery, which is also external to the capacitor and keeps it at a constant potential, $V$. This infinitesimal work amounts to $\delta W_{\rm B}= V \d q$ (we use the symbol $\delta W$ to denote infinitesimal work; mathematically this is an inexact differential, that should not be confused with an infinitesimal variation, such as $\d U$, that is an exact differential), where $q$ stands for electric charge. Since the battery is not included in the system, altogether the work of the force applied to the upper plate and the work of the electric forces are the energy transfers that cause the  system's energy variation. The infinitesimal version of equation (\ref{totale}) is now written as
\beq
\delta W_{\rm ext} = F \d x + V \d q = \d U = \frac{1}{2}V\d q
\label{eq15}
\eeq
(remember that the energy stored in a capacitor is $U= \frac{1}{2}V q$). Equation (\ref{eq15}) leads to
\beq
F \d x = - \frac{1}{2}V\d q\, .
\eeq
Since $\d x >0$, one concludes that $\d q <0$, i.e. in the process there is electric charge flowing out of  the capacitor. Through the action of $\vec F$, external work is performed upon the system  but the capacitor energy ultimately get's reduced  due to the work of the electric forces  responsible for the energy flow from the system. The lower plate is fixed (like particle 1 in section 2) so there is no work performed by the
external force on that plate. Similarly, there is no work done by the internal attractive force exerted on that plate by the upper one.
 Hence, the work of the internal forces on the plates reduces to the work of a force symmetric of $\vec F$ (not represented in figure \ref{fig4n}, but the situation is very similar to the discussion presented in section 2). We denote the work of this force by $\delta W_{\rm int}^{(1)}= -F \, \d x$.
The index (1) is used because there should be another type of work by  internal forces, as we shall see in the sequel,
denoted by $\delta W_{\rm int}^{(2)}$.
Actually, when the electric charges --- the constituents of the system --- flow from/to the capacitor, due to the action of  the battery, they should also experience internal forces.
For the capacitor, its total energy variation, regarded  from ``inside" the system, can be totally assigned to work of internal forces:
\beq
\d U = {1\over 2} V \d q = - F \d x = - \delta W_{\rm int}= - (\delta W_{\rm int}^ {(1)}+ \delta W_{\rm int}^ {(2)})\, .
\eeq

The work of the internal forces encountered in this example deserves a special attention. Actually,
since  $\delta W_{\rm int}=F \d x >0$ and $\delta W_{\rm int}^{(1)}= -F \, \d x$, one concludes that $\delta W_{\rm int}^ {(2)}= 2\, F \d x$. The work $\delta W_{\rm int}^ {(1)}$ is the result of an internal  macroscopic force pointing downwards, applied on the upper plate, but the forces responsible for $\delta W_{\rm int}^ {(2)}$ are not even representable pictorially. This work is ultimately due to the myriad of works performed by each internal force acting on each moving charge that leaves or arrives to the capacitor. The situation is identical to the one found in section 3 for the internal work in the van der Waals gas: the work does exist, but it is not computable {\em ab initio}, i.e. $ W_{\rm int}^ {(2)}$ cannot be calculated, as one does for $W_{\rm int}^ {(1)}$, from equation (\ref{trabalhoi}).

Next, let's discuss the same problem from the point of view of figure \ref{fig4n}-II (we will denote the work and the internal energy with primed symbols). Now the external work transferred to the system is only provided by the force applied to the upper plate of the capacitor,
$\delta W'_{\rm ext} = F \d x$ and the energy balance equation (\ref{totale}) reads
\beq
F \d x =  \d U' = {1\over 2} V \d q +\d U'_\xi \, .
\label{uyt9}
\eeq
In the last expression, the first term represents the energy variation of the capacitor and the second term the chemical energy variation in the battery. The energy variation in the battery is  $\d U'_\xi= -V \d q$ ($\d q$ is the charge leaving the capacitor, so from the point of view of the battery the charge variation is $-\d q$).  Altogether the total energy
variation of the system, capacitor + battery, is $\d U'= -{1\over 2} V \d q$. Of course, $\d U' >0$ and, by looking to the system from ``inside", again we may assign this total energy variation to the work of internal forces, $\d U'=-\delta W_{\rm int}$, with
\beq
 \delta W'_{\rm int}= -F \d x <0 \, .
 \label{negatra}
\eeq
 The energy increase corresponds to a positive external work (\ref{uyt9}) or to the negative work (\ref{negatra}) by the internal forces.
% Now the work of the electric internal forces on the capacitor is equal and opposite to the work of the electric internal forces on the battery.
Now, in case II, the work on the moving charges, mentioned before for  situation I, vanishes because the work of the internal forces on the charges leaving/arriving the capacitor (responsible for decreasing the internal energy of the subsystem capacitor) is equal and opposite of the work of the internal forces on the charges arriving/leaving the battery (responsible for the increase $\d U'_\xi$ of the internal energy of the subsystem battery).
Actually, for the system capacitor + battery, the internal work reduces to the work performed by the internal force, $-\vec F$, exerted on the upper plate.

The physical results for I and II are exactly the same, as they should, a point that, nevertheless, it is worth to emphasize.
The situation is even amazing: by means of work of an external force, one transfers energy to the capacitor but it reduces its energy because the process ultimately leads to an energy transfer to the battery. By moving away the capacitor plates we are charging the battery!

Usually, in textbooks, the example presented in this section is discussed aiming at obtaining the expression of the force, which depends on the plate separation, and is given by $F={\epsilon_0V^2 A\over 2 x^2} $.
Here, we have centred the discussion on  energetic issues, to illustrate that energy variations of the system (irrespective of the way the system is defined) can be regarded as a consequence of the work of internal forces.

A simpler example would consist of a charged isolated capacitor submitted to an external force, similar to $\vec F$, pulling the upper plate (with the lower one fixed). The discussion of this example follows {\em pari passu}
the arguments presented in section 2 for the two particles and the conclusion will be the same: the  internal force is symmetric of  $\vec F$, its displacement is the same of the external force, hence $W_{\rm int}= -W _ {\rm ext}$. Again, one concludes that $\Delta U = W_{\rm ext}= -W _ {\rm int}$.

{\color{black} With the examples studied in this section, once again we put in evidence that, although the internal work might not be computed from equation (\ref{trabalhoi}), it plays a central role if one wants to understand, from an internal point of view, the energy variation of a system,  which of course results from an energetic interaction  with the surrounding.}

\section{A note on double counting and conclusions}

We have discussed a few examples that illustrate the role of the work of the internal forces in the energy variation of a system. We noted in the Introduction that equation (\ref{decomp}) is not an
exhaustive one. However, one should be careful if we add a term of the type $-\sum_j W_{{\rm int,}j}$ to equation (\ref{decomp}) because of double counting issues. Sometimes, the work of the internal forces only intermediates the variation of one type of
internal energy into another type and, in this case, the previous summation still applies but it must include two terms, equal in magnitude with opposite signs, that cancel out. This is, for instance, the case of the automobile studied in \cite{guemez13b} where the variation of chemical energy leads, among others, to rotational energy
in the wheels, by means of internal work. Also in the walking process studied in \cite{guemez13c}, the accelerating phase with the increase of the  centre of mass velocity is ultimately due to
the internal energy decrease originated by biochemical reactions, $\Delta U_\xi$, but the process is intermediated by work of internal forces: $\Delta U=\Delta U_\xi = -W_{\rm int}= -\Delta K_{\rm cm}$.
Adding up $\Delta U_\xi$  and $-W_{\rm int}$ would be an absurd (because of double counting).

The examples studied in this paper, of mechanical, thermodynamical and electromagnetic nature, {\color{black} in our opinion} help to understand the importance of the internal forces
whose role sometimes is underrated by instructors, because those forces are  regarded only from the Newton's third law perspective.


\begin{thebibliography}{99}
\bibitem{papersAJP} W. H. Bernard, {\em Internal work: A misinterpretation}, Am. J. Phys. {\bf 52}, 253-254 (1984);
H. R. Kemp, {\em Internal work: A thermodynamic treatment}, Am. J. Phys. {\bf 53}, 1008 (1984)
\bibitem{guemez13} J. G\"u\'emez, M. Fiolhais, {\em From mechanics to thermodynamics: analysis of selected examples}, Eur. J. Phys.  {\bf 34} 345-357 (2013)
\bibitem{guemez14} J. G\"u\'emez, M. Fiolhais, {\em Thermodynamics in rotating systems -- analysis of selected examples}, Eur. J. Phys.  {\bf 35}  (2014) 015013  (14pp)
\bibitem{afinn} M. Alonso,  E. J. Finn, {\em On the notion of internal energy},  Phys. Educ. {\bf 32} 256-264 (1997)
\bibitem{zemansky} M. W. Zemansky, R. Dittman {\em Heat and Thermodynamics - an intermediate textbook}, 7th edition, McGraw-Hill, Inc., New York (1997)
\bibitem{penchina78} C  M  Penchina, {\it Pseudowork-energy principle},    Am. J. Phys. {\bf 46}, 295-296 (1978)
\bibitem{sherwood83} B  A  Sherwood, {\it Pseudowork and real work},  Am. J. Phys. {\bf 51}, 597-602 (1983)
\bibitem{ray06} S. Ray, J Shamanna {\em On virtual displacement and virtual work in Lagrangian dynamics},   Eur. J. Phys. {\bf 27} 311-330 (2006)
\bibitem{wangsness86} R. K. Wangsness {\em Electromagnetic Fields}, 2nd Edition, John Wiley and Sons, New Jersey, (1986)
\bibitem{guemez13b} J. G\"u\'emez, M. Fiolhais, {\em Forces on wheels and fuel consumption in cars},  Eur. J. Phys.  {\bf 34}  1005–1013 (2013)
\bibitem{guemez13c} J. G\"u\'emez, M. Fiolhais, {\em The physics of a walking robot}, Phys. Educ. {\bf 48} 455-458 (2013)
\end{thebibliography}
\end{document}